\documentclass[aps,pra,twocolumn,superscriptaddress,showpacs,amsmath,amssymb,nofootinbib,longbibliography]{revtex4-1}
\usepackage{hyperref,xcolor,graphicx,multirow,tabularx}

\newcommand{\vout}{V_\mathrm{OUT}}
\newcommand{\vin}{V_\mathrm{IN}}
\newcommand{\figA}{
\begin{figure*}
\includegraphics[width=\columnwidth]{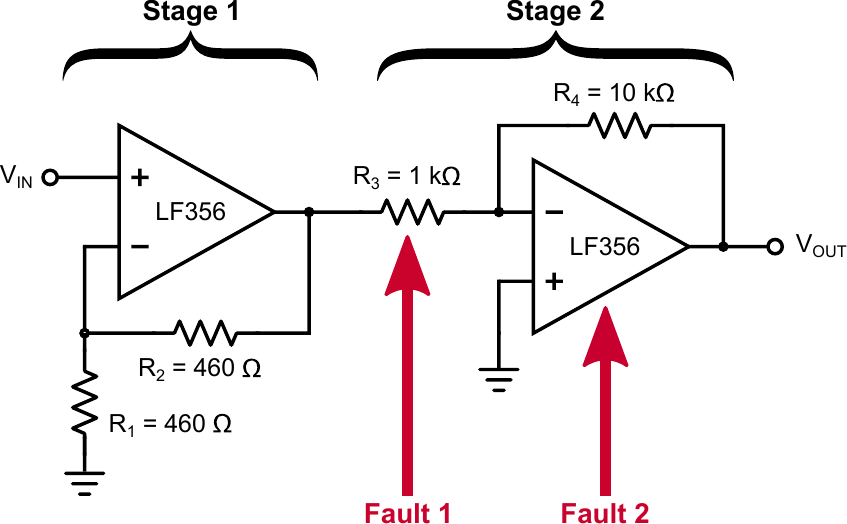} \hfill \includegraphics[width=\columnwidth]{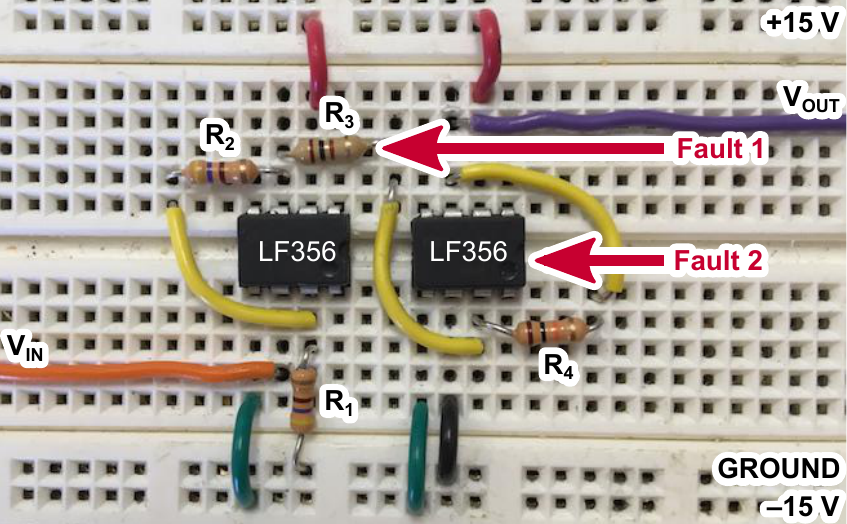}
\caption{\label{fig:schematic}(Left) Annotated schematic diagram for the inverting cascade amplifier, with design elements highlighted. Two stages were connected in series: the first stage, consisting of the leftmost op-amp and resistors $R_1$ and $R_2$, was a non-inverting amplifier with a gain of 2; the second stage, consisting of the rightmost op-amp and resistors $R_3$ and $R_4$, was an inverting amplifier with a gain of $-10$. The handout given to students did not include labels for stages or faults. (Right) Annotated photograph of the physical circuit. The three shown power rails were connected to $\pm15$~V and ground; wires connecting the circuit to power rails are not labeled. The leftmost LF365 op-amp is part of stage~1, and the rightmost is part of stage~2.
}
\end{figure*}
}
\newcommand{\figB}{
\begin{figure*}
\includegraphics[width=\textwidth]{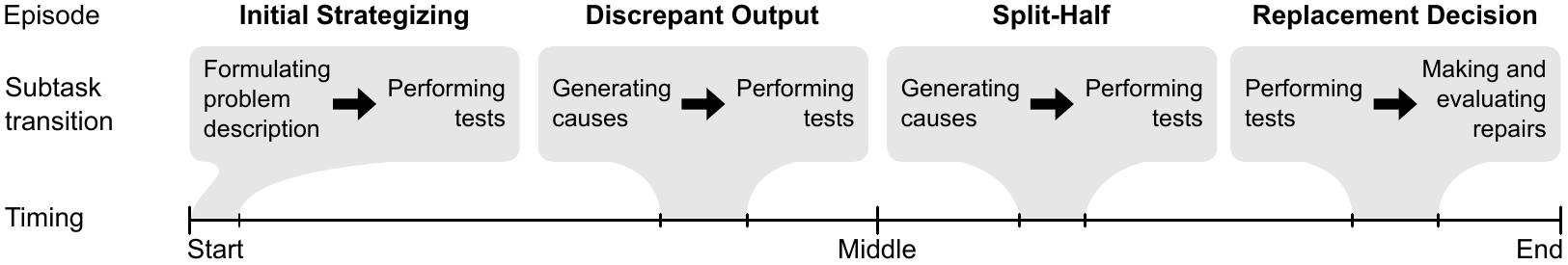}
\caption{\label{fig:episodes}Representation of the ordering and typical duration of episodes selected for analysis. In all interviews, episodes occurred in the order shown here. In half the interviews, either the split-half episode or the replacement decision episode was missing; in these cases, the other three episodes still occurred in the order shown here. Across all interviews, all four episodes together accounted for about 20\% of the total time spent troubleshooting. On average, initial strategizing episodes had the shortest duration. Students typically spent 20 to 45 min troubleshooting the circuit.}
\end{figure*}
}
\newcommand{\tabA}{
\begin{table}
\caption{\label{tab:metacog}Metacognitive moves. Shown are the numbers of groups engaging in dialogue in which at least one conversational turn was coded as a metacognitive move. Results are broken down by type and subtype of metacognitive move. IS episodes were identified in 8 groups, DO in 8 groups, SH in 4 groups, and RD in 7 groups.
}
\begin{ruledtabular}
\begin{tabular}{llcccc}
 & & \multicolumn{4}{c}{Episode category} \\ \cline{3-6}
Type & Subtype & IS & DO & SH & RD \\ \hline
New idea
& Suggest approach & 8 & 8 & 4 & 7 \\
& Articulate observation & 6 & 8 & 2 & 6 \\
& Articulate fact & 6 & 8 & 4 & 3 \\
& Articulate prediction & 3 & 4 & 2 & 4 \\
& Suggest explanation & 0 & 3 & 2 & 3 \\ 
Assessment
& Assess result & 3 & 8 & 3 & 7 \\ 
& Assess own understanding & 3 & 2 & 2 & 3 \\
& Assess approach & 1 & 4 & 0 & 0 
\end{tabular}
\end{ruledtabular}
\end{table}
}
\newcommand{\tabB}{
\begin{table}
\caption{\label{tab:transact}Transactive moves. Shown are the numbers of groups engaging in dialogue in which at least one conversational turn was coded as a transactive move. Results are broken down by type and subtype of transactive move.
}
\begin{ruledtabular}
\begin{tabular}{llcccc}
 & & \multicolumn{4}{c}{Episode category} \\ \cline{3-6}
Type & Subtype & IS & DO & SH & RD \\  \hline
Self-disclosure
& Clarify  & 6 & 8 & 4 & 7 \\
& Justify  & 3 & 4 & 2 & 5 \\ 
Other-monitoring
& Monitor ideas & 5 & 8 & 4 & 7 \\
& Monitor actions & 2 & 4 & 1 & 4 \\ 
Feedback request & & 7 & 6 & 2 & 6 \\
Idea request & & 2 & 1 & 0 & 2
\end{tabular}
\end{ruledtabular}
\end{table}
}
\newcommand{\tabC}{
\begin{table}
\caption{\label{tab:clusters} {Clusters. Shown are the numbers of groups whose dialogue yielded at least one cluster, by theme.}}
\begin{ruledtabular}
\begin{tabular}{lcccc}
 & \multicolumn{4}{c}{Episode category} \\ \cline{2-5}
 Conversational theme & IS & DO & SH & RD \\ \hline
 Collective strategizing	& 3 & 4 & 0 & 1\\
 Shared understanding	& 1 & 4 & 4 & 2 \\
 Neither theme			& 0 & 1 & 1 & 1 
\end{tabular}
\end{ruledtabular}
\end{table}
}

\begin{document}


\title{Investigating the role of socially mediated metacognition during collaborative troubleshooting of electric circuits}

\author{Kevin L. Van De Bogart}
\email{kevin.vandebogart@maine.edu}
\affiliation{Department of Physics and Astronomy, University of Maine, Orono, ME 04469, USA}

\author{Dimitri R. Dounas-Frazer}
\affiliation{Department of Physics, University of Colorado Boulder, Boulder, CO 80309, USA}

\author{H. J. Lewandowski}
\affiliation{Department of Physics, University of Colorado Boulder, Boulder, CO 80309, USA}
\affiliation{JILA, National Institute of Standards and Technology and University of Colorado Boulder, Boulder, CO 80309, USA}

\author{MacKenzie R. Stetzer}
\affiliation{Department of Physics and Astronomy, University of Maine, Orono, ME 04469, USA}
\affiliation{Maine Center for Research in STEM Education, University of Maine, Orono, ME 04469, USA}

\date{\today}

\begin{abstract}
{Developing students' ability to troubleshoot is an important learning outcome for many undergraduate physics lab courses, especially electronics courses. In other work, metacognition has been identified as an important feature of troubleshooting. However, that work has focused primarily on \emph{individual} students' metacognitive processes or troubleshooting abilities. In contrast, electronics courses often require students to work in \emph{pairs}, and hence students' in-class experiences likely have significant social dimensions that are not well understood. In this work, we use an existing framework for socially mediated metacognition to analyze audiovisual data from think-aloud activities in which eight pairs of students from two institutions attempted to diagnose and repair a malfunctioning electric circuit. In doing so, we provide insight into some of the social metacognitive dynamics that arise during collaborative troubleshooting. We find that students engaged in socially mediated metacognition at multiple key transitions during the troubleshooting process. Reciprocated metacognitive dialogue arose when students were collectively strategizing about which measurements to perform, or reaching a shared understanding of the circuit's behavior. In addition to elaborating upon these findings, we discuss implications for instruction, and we identify areas for potential future investigation.}
\end{abstract}

\maketitle


\section{Introduction}

Many undergraduate electronics lab courses are characterized by apprenticeship-style learning environments in which instructors coach pairs of students as they (the students) collaboratively design, build, and troubleshoot electric circuits~\cite{Dounas-Frazer2017}. In particular, while the ability to troubleshoot is an important student learning outcome for undergraduate labs in general~\cite{JTUPP2016,AAPT2015}, it is an especially important goal for electronics courses since the circuits that students are required to build often do not initially work as expected~\cite{Dounas-Frazer2016b}. In most lab courses (and throughout this work), troubleshooting is defined as the process of diagnosing and repairing a malfunctioning apparatus in order to bring its actual performance into alignment with its expected performance. In this sense, troubleshooting is a type of problem solving where the solution state is known, but the nature of the problem is not~\cite{Jonassen2006}. {Thus, many electronics courses regularly engage students in solving experimental physics problems. Moreover, students typically work in groups to collaboratively solve problems that inevitably arise.} Pairwise troubleshooting is a social aspect of learning that is one of the defining features of electronics courses. In the present work, we provide insight into some of the social dynamics that arise when pairs of students work together to troubleshoot a malfunctioning circuit. In particular, we focus on students' {social mediation of metacognition} during multiple key transitions in the troubleshooting process.

Troubleshooting is a nonlinear and iterative problem solving task that involves frequent transitions between multiple subtasks (e.g., generating causal hypotheses and enacting potential repairs). Successful troubleshooting requires more than just sufficient content knowledge; troubleshooters also need to know how to use test and measurement equipment, and they must be able to strategically prioritize which measurements to make and in what order~\cite{Jonassen2006,Schaafstal2000,Perez1991}. Metacognition---or ``thinking about one's own thinking"---has been shown to be an integral component of similarly complex problem solving scenarios in a wide range of mathematics and science contexts~\cite{Schoenfeld1985,Zohar2012,Veenman2012}, {including some aspects of problem solving in introductory physics labs~\cite{LippmannKung2007}}. Hence, it is likely that metacognition also plays an important role in troubleshooting. For example, to diagnose a problem, troubleshooters must continually monitor their progress, evaluate new information, and incorporate that information into their decisions about how to proceed. Along these lines, in a review of research on teaching troubleshooting, Perez~\cite{Perez1991} identified the development, planning, and evaluation of strategies for isolating faults as an example of metacognition specifically relevant to troubleshooting. However, research on the relationship between metacognition and troubleshooting is sparse (see, e.g., Refs.~\cite{vanGog2005a,Pate2011}), and we are unaware of work that explores this relationship in the context of upper-division physics lab courses.

{Some studies have explored metacognition that occurs during small group problem solving in physics~\cite{LippmannKung2007} and mathematics~\cite{Schoenfeld1989,Goos2002} learning environments.} For example, Goos et al.~\cite{Goos2002} reconceptualized metacognition as a social practice in their foundational work on the phenomenon of \emph{socially mediated metacognition (SMM)}, i.e., the process through which metacognition is mediated by collaborative peer interaction. Their findings, which are situated in the context of high school mathematics problem solving, suggest that productive metacognitive decisions can be facilitated by discussions through which students make their thinking ``public and open to critical scrutiny" (p.~219). As lab instructors and education researchers involved with teaching and learning in electronics courses, {we were interested in investigating whether} similar social dynamics might inform the collaborative troubleshooting that takes place when students work together to design, build, and repair circuits.

In the present work, we describe an exploratory qualitative study in which we adapt and apply Goos et al.'s SMM framework to investigate the social metacognitive dynamics that arise as pairs of students attempt to repair a malfunctioning electric circuit. We report on think-aloud interviews with eight pairs of students at two institutions. Preliminary results from this study have been reported elsewhere~\cite{VanDeBogart2015}; here we provide a more comprehensive analysis. This study was designed to address two research questions:
\begin{enumerate}
\item[RQ1.] Do pairs of students engage in socially mediated metacognition while troubleshooting a circuit?
\item[RQ2.] What role does socially mediated metacognition play during the troubleshooting process?
\end{enumerate}
This work not only helps clarify the relationship between metacognition and troubleshooting, it also represents an important step toward understanding the interplay of cognitive, metacognitive, and social aspects of learning in upper-division lab courses.

This paper is organized as follows. In Sec.~\ref{sec:background}, we highlight relevant background literature on troubleshooting and metacognition. We describe the theoretical frameworks underlying our investigation in Sec.~\ref{sec:frameworks}, our data collection and analysis methods in Sec.~\ref{sec:methods}, and the results from our analyses in Sec.~\ref{sec:results}. In Sec.~\ref{sec:discussion}, we discuss our findings and identify implications for research and teaching. Finally, in Sec.~\ref{sec:conclusion}, we provide a brief summary our study.


\section{Background}\label{sec:background}
Our work resides in the intersection of three overlapping educational domains: electronics, troubleshooting, and metacognition. In order to situate our study in these broader contexts, we provide a brief summary of relevant research in these three areas, with a particular emphasis on research related to physics education.

Within the physics education literature, there is a broad spectrum of research on electronics at both introductory and upper-division levels. Some of this work has focused on the design or evaluation of electronics courses~\cite{Coppens2016b,Halstead2016,Lewandowski2015,Mazzolini2011,Getty2009,Shaffer1992}, while other work has focused on student understanding of circuits, circuit components, or related concepts~\cite{Papanikolaou2015,Stetzer2013,Coppens2012,Kautz2011,Engelhardt2004,McDermott1992,*McDermott1993}. Recently, two studies have explored instructor perspectives about teaching upper-division electronics lab courses: Coppens et al.~\cite{Coppens2016a} surveyed students and instructors at multiple Belgian colleges about  learning goals for electronics labs, and Dounas-Frazer and Lewandowski~\cite{Dounas-Frazer2017,Dounas-Frazer2016b} conducted an inter-institutional interview study with electronics instructors across the United States. The latter study focused on instructors' perceptions and practices related to teaching students how to troubleshoot.

Dounas-Frazer and Lewandowski~\cite{Dounas-Frazer2017} showed that, for the instructors in their study, developing students' ability to troubleshoot was a central learning goal of electronics courses, in part because it makes students ``useful in the lab" (p.~6). This finding complements a result from a related study: interviews with physics graduate students at a large research university suggest that knowing how to fix analog electronics is an important aspect of graduate-level experimental physics research~\cite{Pollard2014}. Nevertheless, there is a dearth of research on physics students' troubleshooting abilities in undergraduate electronics environments---one exception being our own previous work on students' use of model-based reasoning while troubleshooting an electric circuit~\cite{Dounas-Frazer2016a,Dounas-Frazer2015}. In that work, we showed that modeling and troubleshooting are overlapping processes, and argued that ``courses designed to develop students' ability to troubleshoot should also emphasize students' ability to model physical systems"~\cite{Dounas-Frazer2016a} (p.~18). 

Troubleshooting is common to numerous professional contexts, such as diagnosing illnesses and debugging computer programs. Accordingly, there is a large body of literature on troubleshooting across disciplines (see Refs.~\cite{Jonassen2006,Schaafstal2000,Perez1991} for overviews). {In the domain of electric circuits and other electrical systems, common research foci related to troubleshooting instruction} include developing and evaluating training programs and educational interventions for high school students~\cite{vanGog2008, vanGog2006, Kester2006, Kester2004} and students in technical fields~\cite{Ross2009,Johnson1999,MacPherson1998,Estrada2012,deCroock1998,Johnson1993}. Given our interest in the role of metacognition in troubleshooting circuits, one study is particularly relevant: van Gog et al.~\cite{vanGog2005a} observed that high school students tended to make ongoing assessments of their actions when troubleshooting a simulated circuit, and suggested this may be related to their metacognitive knowledge. However, no framework for metacognition was used in their analysis.

An extensive discussion of current research on metacognition can be found in Ref.~\cite{Zohar2012}. Although social aspects of metacognition have not been a major research focus in the physics education literature, there have been some studies along these lines (see, for example, a recent study that focuses on students' spontaneous metacognitive talk~\cite{Sayre2015}). In the context of introductory physics labs, Lippmann Kung and Linder~\cite{LippmannKung2007} found that groups of students regularly verbalized metacognitive statements, but that ``more critical is \emph{how} students react to this metacognition" (p.~54; italics in original). In their study, students' metacognition did not always result in modified student approaches to lab activities. Accordingly, Lippmann Kung and Linder emphasized the importance of focusing on students' reactions to metacognition, as we do here.

The frameworks that directly informed our study focus mostly on metacognitive regulation of either an individual's thinking~\cite{Schoenfeld1987} or a group's thinking~\cite{Goos2002}. Schoenfeld~\cite{Schoenfeld1987} examined the role of self-regulation in undergraduate mathematics problem solving. His work focused on the task of managing oneself during the problem solving process, including the need for verifying one's understanding of a problem, planning how to solve the problem, monitoring the effectiveness of a solution, and deciding how to allocate time~\cite{Schoenfeld1987}. The need for a social framework for metacognition arose from an effort by Goos~\cite{Goos1994} to study the metacognitive strategies employed by pairs of mathematics students working on introductory physics problems. Goos initially employed a methodology similar to that used by Schoenfeld~\cite{Schoenfeld1985}, segmenting and characterizing time in interviews according to when specific behaviors were demonstrated. However, Goos found that, while this approach captured macroscopic features of problem-solving, another framework was needed to describe the nature of the interactions between individuals~\cite{Goos1994}.

Using ideas from Vygotsky's work~\cite{Vygotsky1980}, Goos and Galbraith~\cite{Goos1996} expected that, through collaboration, students would complement and enhance one another's knowledge, {establishing a zone of proximal development} and thus resulting in collaborative performance exceeding that of either student individually.  Goos and Galbraith noted that both the quality of metacognitive decision-making and the nature of the social interactions between subjects significantly influenced the outcomes of problem solving activities.  To further explore the latter interaction, Goos et al.~\cite{Goos2002} developed the socially mediated metacognition framework, which we describe in detail in the following section.


\section{Theoretical frameworks}\label{sec:frameworks}

Throughout this work, we define troubleshooting as the process of diagnosing and repairing a malfunctioning apparatus. Our goal is to identify and describe examples of how students socially mediate their metacognition while collaboratively troubleshooting an electric circuit. As a result, the theoretical grounding of this work is rooted in two  complementary perspectives: a cognitive task analysis of troubleshooting~\cite{Jonassen2006,Schaafstal2000,Johnson1988}, and the socially mediated metacognition framework, which describes the metacognitive dynamics that arise among students during group problem solving processes~\cite{Goos2002}. In this section, we describe and synthesize each of these theoretical perspectives. When appropriate, we use examples from electronics to help illustrate {these} ideas.


\subsection{Troubleshooting as a cognitive task}\label{sec:CTA}

Troubleshooting {typically} requires a high level of cognitive activity: making decisions and judgments, paying attention to details of models and apparatuses, analyzing and interpreting the results of measurements, and so on. Hence, troubleshooting is often interpreted as a cognitive task. Corresponding cognitive task analyses typically describe the subtasks and types of knowledge associated with troubleshooting~\cite{Jonassen2006,Schaafstal2000,Johnson1988}. Indeed, we have relied on these aspects of troubleshooting in other studies of electric circuits~\cite{Dounas-Frazer2016a} and electronics instruction~\cite{Dounas-Frazer2017}. In this section, we summarize the cognitive elements of troubleshooting that are relevant for the present work.

The troubleshooting process can be subdivided into four subtasks: \emph{formulating the problem description}, \emph{generating causes}, \emph{performing tests}, and \emph{making and evaluating repairs}~\cite{Schaafstal2000}. Formulating the problem description refers to the initial phase of troubleshooting, during which the troubleshooter performs preliminary inspections and measurements in order to determine which portions of the system work as expected and which do not. Generating causes involves forming causal hypotheses that may explain the circuit's malfunctioning behavior. Hypotheses are tested by performing diagnostic measurements with oscilloscopes, multimeters, or other devices. Last, repairs to a circuit include rewiring erroneous connections, replacing faulty components, and other revisions to the apparatus. The performance of the revised circuit must be evaluated in order to determine whether the troubleshooting process is complete. If the circuit functions as expected, the troubleshooting process comes to a stop. Troubleshooters often engage in these subtasks in nonlinear and recursive ways. For example, depending on the outcome of diagnostic tests of a causal hypothesis, a troubleshooter may either generate additional causes (if the original hypothesis was incorrect) or enact a repair (if it was correct).

Troubleshooting is facilitated by multiple types of knowledge, including \emph{domain}, \emph{system}, \emph{strategic}, and \emph{metacognitive} knowledge~\cite{Jonassen2006,Pate2011,vanGog2005a}. Domain knowledge refers to the theories and principles that underlie electric circuits, including models like Kirchhoff's laws and concepts like equipotential surfaces. System knowledge refers to the structure and function of component blocks, and how they impact electron flow and voltage drops across interconnected circuit subsystems. Strategic knowledge is knowledge about how to act; it consists of heuristic techniques and methodical approaches to troubleshooting the system. One example of a strategy that is used by many students in our study is the \emph{split-half strategy}. The split-half strategy reduces the problem space through a binary search; the circuit is divided into two subsystems, and diagnostic tests are performed in order to isolate one of the two subsystems as the source of fault. {Last, metacognitive knowledge includes knowledge about which strategy to use, when to use it, and why.}

{Metacognitive knowledge is only one aspect of metacognition; metacognition also consist of metacognitive skills, i.e., the ability to control one's own problem-solving approaches~\cite{Veenman2012}. For example, Perez~\cite{Perez1991} defines metacognitive processes as ``the knowledge \emph{and} control a troubleshooter has over his or her own thinking and activities" (p.~121; emphasis added). Two categories of metacognitive skills are \emph{self-monitoring} and \emph{self-regulation}~\cite{Perez1991,Jacobs1987}.} Self-monitoring includes not only understanding and communicating one's own thought processes~\cite{Jacobs1987}, but also being aware of the strategies and resources needed to troubleshoot effectively~\cite{Perez1991}. Self-regulation---which Schoenfeld~\cite{Schoenfeld1987} argued is particularly relevant in the context of mathematical problem solving---involves consideration of how to perform long tasks and ensure their successful completion~\cite{Jacobs1987,Perez1991}; along these lines, van Gog et al.~\cite{vanGog2005a} argue that metacognitive knowledge ``is used to monitor [the troubleshooting process] by keeping track of the progress toward the goal state" (p.~237).

Our goal is to explore how metacognition is mediated by interaction among pairs of students collaboratively troubleshooting a circuit. Because social dynamics are omitted from the cognitive task analyses of troubleshooting with which we are familiar~\cite{Jonassen2006,Schaafstal2000,Johnson1988}, our study relies on the socially mediated metacognition framework.


\subsection{Socially mediated metacognition}\label{sec:SMM}

{Here, we provide an overview of socially mediated metacognition. We draw on the work of Goos et al.~\cite{Goos2002}, who developed the SMM framework in order to capture instances where metacognition is mediated by peer interaction. In  this framework, mediation of metacognition occurs through discussion among students. Two grain sizes of discussion are relevant: (i)~individual conversational turns, called ``moves;" and (ii)~exchanges between students that consist of multiple successive turns of dialogue, referred to as ``clusters."}

{A single conversational turn may be characterized as a move that has a metacognitive function and/or a transactive quality; we refer to these as \emph{metacognitive moves} and \emph{transactive moves}, respectively.} The SMM framework distinguishes between two types of metacognitive moves: \emph{new ideas} and \emph{assessments}. A student contributes a new idea to the discussion when they {introduce} new and potentially useful information, or when they propose an alternative problem solving approach. A variety of assessments constitute metacognitive moves: whether a strategy is appropriate and being executed with care, whether a result is accurate and sensible, or whether one's own knowledge and understanding are sufficient. Transactive moves are interpersonal by definition, and are meant to characterize how students interact with one another's ideas. Drawing from work on peer collaboration~\cite{Kruger1993}, Goos et al.~\cite{Goos2002} identify three types of transactive moves in the SMM framework: \emph{self-disclosure}, \emph{other-monitoring}, and \emph{feedback requests}. Students making such moves seek to clarify, elaborate, or justify their own reasoning (self-disclosure) or that of their partners (other-monitoring). They may also solicit critiques of their own ideas (feedback requests). A given conversational turn can be metacognitive, transactive, both, or neither.

The extent to which transactive or metacognitive moves contribute to mediation of metacognition depends on the details of the discussion in which they occur. Within the SMM framework, the concepts of \emph{metacognitive nodes} and \emph{transactive clusters} help characterize the degree to which particular moves are connected through discussion about a common theme. Metacognitive nodes refer to instances of dialogue where a metacognitive move is either prompted by, or results in, a transactive move. Metacognitve and transactive moves that comprise a node are said to be ``connected." Transactive clusters arise when a metacognitive move is connected to more than one transactive move.

When describing transactive clusters, Goos et al. suggest that such ``discussion \emph{around}, and generated by, individual metacognitive acts is crucial to the success of {the mathematical enterprise}" (p.~213; italics in original). Indeed, Goos et al. found significantly higher rates of transactive clusters among student groups that were successful at collaborative problem solving compared to those that were not. Transactive interactions around metacognitive decisions enabled student groups to notice errors in their reasoning and endorse fruitful problem solving strategies, ultimately facilitating successful navigation of challenges that arose during the problem solving process. Thus, in the SMM framework, the most impactful mediation of metacognition occurs through transactive clusters. 

{Goos et al. originally developed the SMM framework to document metacognition that stems from group collaboration in mathematics. The framework has since proven to be flexible enough to be adapted to other contexts: middle school computer programming~\cite{Werner2009} and an educational psychology course for teachers~\cite{Siegel2012}. It also informed the work of Lippmann Kung and Linder~\cite{LippmannKung2007}, who focused on metacognition among groups of students in introductory physics labs. In this work, we map the SMM framework to yet another context: the cognitive task of troubleshooting.}


\subsection{Synthesizing the frameworks}\label{sec:synth}

The cognitive task analysis of troubleshooting and the socially mediated metacognition framework each provide a distinct lens through which to understand collaborative troubleshooting of electric circuits. Nevertheless, these lenses are connected. In this section, we highlight synergies between the two perspectives by describing how SMM may arise during different troubleshooting subtasks, and how different types of troubleshooting knowledge may inform metacognitive and transactive moves.

{Any time a measurement is performed on a malfunctioning circuit, metacognitive moves through which one student brings to light new information may occur.} For example, when formulating the problem description, a student may verbalize the results of their initial visual inspection of the circuit. Similarly, they may contribute new information by announcing the results of a diagnostic or evaluative measurement performed during the testing or repair phase of troubleshooting. Any time new measurements or observations are performed, a student may also assess whether that information is sensible based on their understanding of the expected function of the circuit. Such assessments are grounded in the student's domain and system knowledge, which inform expectations about the behavior of a functional circuit.

Other types of metacognitive moves may arise when generating causes. For example, based on previous tests or visual inspections, a student may propose new explanations for the observed behavior of the circuit. Alternatively, after assessing their own domain and system knowledge, the student may acknowledge that they do not know what to make of the available evidence, and hence cannot hypothesize about what may be causing the malfunction. During the testing phase, metacognitive moves include assessing whether the current strategy is appropriate or proposing a new strategy altogether. Such assessments and proposals rely on students' strategic knowledge.

Transactive moves could likewise occur during any troubleshooting subtask. A student may feel the need to justify their reasoning to their partner when proposing a new hypothesis to explain the circuit's behavior, a new strategy for performing tests, or a new idea about how to repair the circuit. Alternatively, a student may solicit feedback from their partner because they lack conviction in their proposal, which they may frame as speculative. In response, their partner may ask follow-up questions in order to better understand what was proposed, and why.

Because metacognitive and transactive moves likely occur in all troubleshooting subtasks, it is reasonable to expect that metacognitive nodes and transactive clusters also arise throughout the process. {In particular, nodes and clusters may arise when students must collaboratively decide what to do next (e.g., which measurement to perform or which component to replace). As we will show, such decisions occur during transitions between troubleshooting subtasks.} In the present work, we investigate whether and how SMM arises as pairs of students transition from one troubleshooting subtask to the next. We use the cognitive task analysis of troubleshooting to help us identify key episodes during students' troubleshooting processes, and we use the SMM framework to capture students' fine-grained metacognitive behaviors as they work together to repair the circuit.


\section{Methods}\label{sec:methods}

{To characterize the role of socially mediated metacognition in troubleshooting, we conducted an exploratory and qualitative study. We carried out interviews with eight pairs of physics students who were asked to diagnose and repair a malfunctioning electric circuit while thinking aloud.} We have previously used data from the participants in this study to explore connections between troubleshooting and students' model-based reasoning~\cite{Dounas-Frazer2016a,Dounas-Frazer2015}. The present work focuses on social metacognitive dynamics that were beyond the scope of our prior efforts. Elsewhere, we have reported a preliminary analysis of students' socially mediated metacognition~\cite{VanDeBogart2015}. Here, we expand on that work by providing a more detailed analysis that aims to answer our research questions, RQ1 and RQ2: do pairs of students engage in SMM while troubleshooting a circuit, and what role does SMM play during the troubleshooting process?

In this section, we describe the study participants, design of the troubleshooting activity, think-aloud protocol, data analysis methodology, and coding scheme that we used for our study.

\figA


\subsection{Participants and course context}

A detailed description of the participants and course context is presented in Ref.~\cite{Dounas-Frazer2016a}. We present a more concise version here. Participants in this study were physics majors at either the University of Colorado Boulder (CU) or the University of Maine (UM). Eight pairs of students, four from CU and four from UM, were interviewed for this study, for a total of 16 unique participants. Commensurate with student demographics in the undergraduate programs at both institutions, participants were predominantly white men. All participants were enrolled in an upper-division electronics course during Fall 2014. That semester, the third and fourth authors taught the electronics courses at CU and UM, respectively, and the first author was a teaching assistant in the UM course.

The electronics courses at CU and UM are required for physics majors, and are typically taken in the third year of instruction. The courses are each one semester in length and cover a similar spectrum of topics, with an emphasis on analog components and devices such as diodes, transistors, and op-amps. Both courses meet three times per week: twice for 50-minute lectures, and once for lab (three hours at CU, two at UM). During lab, students work together in pairs to complete guided lab activities. At the time of this study, neither course included lectures on troubleshooting strategies. Consistent with the practices of other electronics instructors~\cite{Dounas-Frazer2017}, formal instruction about troubleshooting took place almost exclusively via apprenticeship-style interactions during lab activities.

In this study, participants were tasked with troubleshooting a circuit consisting of two operational amplifiers (op-amps). Both the CU and UM courses focus on op-amps and their use in a variety of practical applications. In these courses, students are taught that an op-amp is a high-gain differential amplifier with an inverting input, non-inverting input, a single output, and two power connections. The power connections are typically attached to positive and negative 15~V supplies, often referred to as power rails. Students are taught a first-order model of op-amps in circuits that employ negative feedback. This model describes the functional behavior of op-amps in such circuits via two \emph{golden rules for op-amps}, articulated by Horowitz and Hill~\cite{Horowitz1989} as: ``I. The output attempts to do whatever is necessary to make the voltage difference between the inputs zero," and ``II. The inputs draw no current" (p.~177). When used in conjunction with Kirchhoff's laws, the golden rules are sufficient to predict the behavior of many op-amp circuits, including the circuit used in the present study. The golden rules are explicitly covered in both the CU and UM electronics courses.


\subsection{Data collection}

Participant recruitment took place near the end of the fall semester at CU and during the beginning of the following spring semester at UM. Students were invited to participate in the study via email and in-person requests. Students were allowed to select a partner if they wished. Those who did not do so were paired by the interviewers on the basis of availability. Participants were given small monetary incentives for their time, but involvement was strictly voluntary and no course credit was given in exchange for participation. During the interview, pairs of students were tasked with diagnosing and repairing a malfunctioning circuit while thinking aloud. Here, we describe the circuit design and  interview protocol.


\subsubsection{Research task}

In the interviews, students were asked to troubleshoot the inverting cascade amplifier shown in Fig.~\ref{fig:schematic}. The circuit can be divided into two distinct stages, each of which may be analyzed separately. Stage 1 of the circuit, consisting of the leftmost op-amp and resistors $R_1$ and $R_2$, is a non-inverting amplifier with a gain of $G_1\equiv(1 + R_2 / R_1) $. Stage 2, which consists of the rightmost op-amp and resistors $R_3$ and $R_4$, is an inverting amplifier with a gain of $G_2\equiv-R_4 / R_3$. In a functioning circuit, the output is proportional to the input, with a scaling factor (also called the \emph{transfer function} of the circuit) equal to the product of the gains of each stage: $G=G_1G_2$. Hence the output is given by $\vout = -(1+R_2/R_1)(R_4/R_3)\vin$. 

Given the nominal resistor values in Fig.~\ref{fig:schematic}, the expected gains of the first and second stages are $G_1= 2$ and $G_2=-10$, and the nominal gain of the whole circuit is $G=-20$. Therefore, the amplitude of the output voltage signal is 20 times larger than that of the input. For ac signals, the output is $180^\circ$ out of phase with the input. The output voltage is constrained by the voltages of the power rails such that, in practice, the output voltage must always be slightly lower than the positive rail voltage, and slightly higher than the negative rail voltage. Any input voltages that would cause the output to exceed these limits will result in saturation (i.e., the output voltage will be truncated to within a volt or so of each power rail).

We intentionally built the cascade amplifier so that it would malfunction in a particular way. Two principles informed our design: {first, students should be able to engage in multiple iterations of troubleshooting; second, the split-half strategy should be a viable approach for troubleshooting the circuit}. In accordance with the first principle, we introduced two different faults. In accordance with the second, we located both faults in stage~2. Since the faults affected solely the performance of stage~2, a student could in principle isolate all problematic behavior to that stage alone.

The first fault (``fault~1" in Fig.~\ref{fig:schematic}) was that the resistor $R_3$ was an order of magnitude smaller than its prescribed value. Therefore, the actual gain of stage~2 (and hence of the whole circuit) was larger than the nominal gain by an order of magnitude. On its own, this fault could result in saturation even for a relatively small input voltage. We expected fault~1 to be relatively straightforward to diagnose, as the incorrectly colored bands on the resistor serve as a visible cue, making it possible to diagnose this fault through visual inspection of the circuit. The second fault (``fault~2" in Fig.~\ref{fig:schematic}) was that the op-amp was damaged in such a way that its output voltage was a constant dc voltage approximately equal to the negative rail voltage. The faulty op-amp did not obey the first golden rule.


\subsubsection{Think-aloud interviews}

{We conducted interviews using a think-aloud protocol with pairs of students troubleshooting a pre-constructed circuit. Such protocols, in which subjects are asked to verbalize their thoughts concurrently with their actions, are relatively non-invasive in a paired setting since students frequently clarify their thinking to their partners while justifying differing opinions, etc.~\cite{vanSomeren1994}. The students in this study were accustomed to working in pairs in their electronics labs and, during the interview, they engaged in discussions with one another with minimal outside intervention.}

We designed our study to be both \emph{controlled} and \emph{authentic}. It was controlled in the sense that each pair of students had similar academic preparation and used the same pre-assembled circuit, hence all participants were working from similar initial conditions. The study was authentic in the sense that the interview conditions were as similar to the students' electronics lab experience as possible. Students at each university were presented with a physical setup (i.e., breadboard, components, voltage sources, and measurement equipment) that closely resembled what they had used in their respective courses. All groups had access to a multimeter, oscilloscope, function generator, power supply with variable and fixed voltages, and a suite of replacement components and wires. In addition, when constructing the circuit, we took care to ensure that the wiring was relatively easy to follow. 

The interview itself began when the interviewer presented students with a schematic diagram of the circuit and a datasheet for the op-amp. The interviewer then gave a short introductory prompt to the activity, requesting students to approach the task as if their peers had built the malfunctioning circuit in the lab. (See Ref.~\cite{Dounas-Frazer2016a} for the full text of this prompt.) Students were subsequently presented with the physical circuit and tasked with diagnosing and repairing the circuit. Students were asked to think aloud as they worked, and to act as though the interviewer was not present. If the students were silent for a significant length of time, the interviewer would prompt them to continue speaking. In practice, there was minimal intervention on the part of the interviewer. The activity ended either when the students had completed their repairs, or when roughly one hour had passed. The initial prompt from the interviewer was approximately two minutes in length, and students typically spent between 20 and 45 minutes on the troubleshooting activity. Seven of the eight groups were ultimately able to repair the circuit, while the remaining group ran out of time prior to completing the task. Video and audio data were collected for all interviews, and audio data were used to generate complete transcripts.


\figB

\subsection{Data analysis}

Our study was not designed to compare between pairs of students based on troubleshooting ability or quality of metacognitive discussion. Instead, it was designed to examine the presence and role of socially mediated metacognition during the troubleshooting process. To characterize students' social metacognitive exchanges at different points in the troubleshooting activity, we developed an \emph{a priori} coding scheme based on the SMM framework (Sec.~\ref{sec:SMM}) and applied it to four types of episodes that occurred across multiple pairs of students. We focused on episodes that correspond to transitions between troubleshooting subtasks because we anticipated that such episodes would provide rich examples of social metacognition (Sec.~\ref{sec:synth}). By analyzing multiple pairs' dialogue in a given episode, we hoped to gain insight into the spectrum of moves and clusters that arose as students transitioned from one subtask to the next.

After identifying episodes, we performed within-episode and cross-episode analyses. Within each episode, we performed a line-by-line analysis of the corresponding transcribed dialogue to identify metacognitive and transactive moves; a detailed example of this approach is described elsewhere~\cite{VanDeBogart2015}. We then analyzed successive moves for the presence of nodes and clusters, the latter of which have been associated with particularly impactful examples of metacognition in other group problem solving contexts~\cite{Goos2002}. Thus, within-episode analyses address our first research question (RQ1) by determining whether students engaged in socially mediated metacognition in one or more transitions between troubleshooting subtasks. Across episodes, we looked for emergent patterns among the topics of conversation in which clusters arose. Such cross-episode analysis addresses our second research question (RQ2) by helping us understand, in broad strokes, the ways in which students engaged with one another's ideas.

In this section, we define the four categories of episodes we analyzed, and we describe our within-episode and cross-episode analyses.


\subsubsection{Episode definitions}

Metacognitive moves occurred throughout the duration of all interviews in our study. Students regularly contributed new ideas by announcing the result of a measurement and assessing whether that measurement aligned with their expectations. However, in this work, we are interested in instances where both metacognitive \emph{and} transactive moves  are frequent, and hence dialogue is likely to contain nodes and clusters. As we argued in Sec.~\ref{sec:synth}, we anticipate that nodes and clusters will occur when students transition from one troubleshooting subtask to the next. For example, during transitions, one student may ask the other to justify or clarify their proposed testing strategy, hypothetical cause of malfunction, or suggestion for how to repair the circuit. Such instances would constitute nodes wherein one student monitored the other's new idea. Therefore, in order to constrain our analyses to time intervals in which rich metacognitive dialogue was more likely to occur, we selected four categories of episodes to analyze in detail: \emph{initial strategizing (IS)}, \emph{discrepant output (DO)}, \emph{split-half (SH)}, and \emph{replacement decision (RD)} episodes. In Fig.~\ref{fig:episodes}, we have illustrated how each of these episodes are connected to transitions between troubleshooting subtasks, and how they are related to one another temporally. Here, we provide a definition and rationale for each episode type:

\begin{itemize}
\item[] \textbf{IS:} Initial strategizing episodes captured how students first approached the task. These episodes began once the interviewer finished introducing the problem; they ended when students either began checking the circuit's connectivity or making measurements. IS episodes were expected to be representative of a transition from formulating a description of the problem to performing tests. We identified IS episodes for all eight groups. Most of these episodes lasted from 30 to 60~s, though two IS episodes lasted about 3~min.
\item[] \textbf{DO:}  Discrepant output episodes captured how students responded to a mismatch between the expected output of the circuit and the measured output. These episodes began when students first observed that the output of the entire circuit was a constant dc value; they ended when students enacted a plan to make further measurements. DO episodes were expected to contain a transition from generating causes for their unexpected measurement to performing additional tests. We identified DO episodes for all eight groups, and the duration of these episodes ranged from 1.5 to 3.5~min.
\item[] \textbf{SH:} Split-half episodes captured how students strategized after identifying a working stage in the circuit. These episodes began just after students had eliminated the first stage of the circuit as a source of faults; they ended when students enacted a plan to make further measurements. SH episodes were expected to represent another clear transition from generating causes (necessitated by partially localizing the fault) to performing further tests. Five of the eight groups employed a split-half strategy. Three of these episodes lasted from 1 to 2~min, and one SH episode lasted about 4.5~min.
\item[] \textbf{RD:}  Replacement decision episodes captured how students came to the decision to replace the faulty op-amp. These episodes began just after the completion of the last set of measurements made before students decided to replace the second op-amp; they ended when the op-amp was replaced. We focused on the replacement of the op-amp and not resistor $R_3$ for two reasons: decisions to replace the resistor neither  coincided with extended dialogue between students, nor corresponded to a transition between cognitive tasks. For example, multiple groups replaced the resistor as part of their initial visual inspection of the circuit, a process that continued after the resistor was replaced. Replacement of the op-amp, on the other hand, corresponded to a transition from performing tests to repairing and evaluating the circuit. Seven of the eight groups successfully replaced the faulty op-amp. The duration of most episodes ranged from 1 to 2.5~min, though one episode lasted about 5~min. 
\end{itemize}

The episodes in all four categories occurred in the same order (Fig.~\ref{fig:episodes}), unless a category was not present. The initial strategizing always occurred within the first few minutes of the interview, immediately after the nature of the task had been explained. The discrepant output episodes tended to occur after the first third but before the second half of the interview, while the discussions following a split-half strategy generally occurred in the final third of the interview. Replacement decisions were made in the final quarter of the episode.

{All four episode categories were present in three of the groups. Only three episodes were present in each of the other five groups: one group did not replace the faulty op-amp; one group replaced the op-amp immediately after employing a split-half strategy (this episode was  categorized as an RD episode, not an SH episode); and three groups did not employ a split-half strategy}. In total, we identified 27 unique episodes across the eight participating groups. The cumulative duration of these 27 episodes was approximately one hour, accounting for roughly 20\% of the aggregated interview time for all groups. For all 27 episodes, we coded the corresponding transcripts using the analysis frameworks described in the following sections.


\subsubsection{Within-episode analyses: \emph{A priori} coding scheme}\label{sec:within-episode}

We initially developed operational code definitions based on the SMM framework (Sec.~\ref{sec:SMM}) and the coding scheme used by Goos et al.~\cite{Goos2002}. Our operational definitions were refined through iterative cycles of collaborative coding by the first and second authors, and discussions with the research team as a whole. By ``collaborative coding," we mean that the initial iteration of coding was performed simultaneously by the two coders. During subsequent iterations of coding, the first and second authors first applied codes independently and then resolved all discrepancies through discussion. The final version of our coding scheme deviated in minor ways from the original schemes presented by Goos et al. Here, we first present our final scheme and then note differences from the work of Goos et al.

Our SMM coding scheme involves coding individual conversational turns for their metacognitive function or transactive quality. Metacognitive moves are characterized by statements where one student introduces or assesses ideas. {We identified new ideas and assessments by directly coding for particular types of statements. We used the following scheme (code names in italics):}
\begin{itemize}
\item[] \textbf{New idea:} A student verbally expresses new information that is relevant to the situation. This may occur when the student is \emph{suggesting an approach}; \emph{suggesting an explanation} for the circuit's behavior; \emph{articulating a prediction} of the outcome of an event; \emph{articulating an observation} about the circuit, measurement tools, handout, or datasheet; or \emph{articulating a fact} relevant to the task at hand. Examples include: ``I would start with just checking if the chips are working," ``Maybe this red [wire], the power, is somehow touching the output?," and, ``Oh, hey. Look. [The voltage] stabilized for some reason."
\item[] \textbf{Assessment:} A student attempts to evaluate information. This may occur when
the student is \emph{assessing a result} of a measurement or prediction as reasonable; \emph{assessing an approach} as appropriate; or \emph{assessing their own understanding} of the problem at hand. Examples include: ``The first [stage] is giving us a good voltage,"  ``Yeah, I mean, [replacing the op-amp] will be like the brute force method of making sure it's the right chip," and, ``We have a good output for the first op-amp."
\end{itemize}

Transactive moves are characterized by statements that are verbal requests for interaction with the other participant, which may in turn prompt further dialogue. We coded for instances of self-disclosure, other-monitoring, feedback requests, and idea requests. {To code these instances of speech, we used the following scheme (code names in italics):}
\begin{itemize}
\item[] \textbf{Self-disclosure:} A student \emph{clarifies} or \emph{justifies} their own thinking. Examples include: ``Well, I was just saying that, maybe if these two op-amps are oriented the same way, that the pins for the second one are connected correctly," and, ``It should be a gain of 2 because you have a voltage divider here with the two [resistors]."
\item[] \textbf{Other-monitoring:} One student responds to the other with the aim of critiquing, building upon, or inquiring about what the other student is thinking (\emph{monitoring ideas}) or doing (\emph{monitoring actions}). Examples include:  {``[Your suggested approach] will be like the brute force method,"} and, ``What are you looking for on the oscilloscope?"
\item[] {\textbf{\emph{Feedback request}:} One student asks the other student to critique an idea or approach. For example: ``[The circuit] should be inverting the signal and amplifying it, correct?"}
\item[] {\textbf{\emph{Idea request}:} One student asks the other student to suggest a new idea or approach. For example: ``What's next?"}
\end{itemize}

After we coded the data for the presence of metacognitive and transactive moves, we examined {sequences of} the coded moves for the existence of nodes and clusters in order to systematically capture students' social engagement in one another's ideas. In our SMM scheme, we used the following operational definitions for nodes and clusters:
\begin{itemize}
\item[] \textbf{Node:} Two successive conversational turns, in which either the first turn is a metacognitive move and the second a transactive one, or vice versa. In a given node, the second move must be in reference to the same idea as the first.
\item[] \textbf{Cluster:} A series of two or more overlapping nodes. Nodes are ``overlapping" if a single conversational turn comprises both the second move of the first node and the first move of the second node. Clusters contain at least three unique turns of conversation, each of which functions as a metacognitive and/or transactive move.
\end{itemize}
Nodes and clusters are meant to capture back-and-forth interactions. {We are particularly interested in identifying and characterizing clusters since they constitute a reciprocated verbal exchange between two students.}

As an example of nodes and clusters, consider the following exchange between two students, G1 and G2, that took place after they finished discussing the circuit schematic:\\
\vspace{6pt}\\
\noindent
\begin{tabularx}{\columnwidth}{*{4}{l}X}
1. & & &
G1: & Okay, what is on the sheet is correct. \\ \cline{2-2} 
\multicolumn{1}{l|}{2.} & \multirow{2}{*}{A} & & 
G2: & Alright, the first thing to do is actually check the circuit for all the resistors, and--- \\ \cline{3-3} 
\multicolumn{1}{l|}{3.} & \multicolumn{1}{l|}{} & \multirow{2}{*}{B} &
G1: & Yeah, that's what I'd start with. Check all the resistor values. \\ \cline{2-2}
4. & \multicolumn{1}{l|}{} & &
G2: & Yeah, make sure that they're all connected.\\ \cline{3-3}
5. & & &
G1: &  Make sure that they're all connected. Okay, so we'll turn this on. \\ 
\end{tabularx}
\vspace{6pt}\\
Turn numbers are indicated on the left. Nodes are indicated with square brackets labeled with single letters. In this interaction, turns 2 through 4 form a cluster. Turns 1 and 5 are included to demonstrate the boundaries of the cluster. Whereas diagnostic approaches are the topic of conversation during the cluster, G1 is focused on the datasheet in turn 1 and shifts his attention to a piece of equipment in turn 5. Hence, turns 1 and 5 are not included in the cluster. Within the cluster, G2 suggested a general approach (``check \ldots\ all the resistors"), G1 endorsed and built upon that suggestion by describing a more specific approach (``check all the resistor values"), and, finally, G2 responded to G1 by outlining another approach (``make sure that they're all connected"). Turns 2 and 3 form node A, wherein G2's metacognitive move (new idea) resulted in a move by G1 that was both transactive (other-monitoring) and metacognitive (new idea). Next, turns 3 and 4 form node B, in which G1's transactive and metacognitive move was followed by a metacognitive move by G2. Because nodes A and B both have turn 3 in common, they overlap to form a cluster. Although G1 repeated G2's new idea in turn 5, G1 was not obviously endorsing or critiquing that idea. Therefore, turn 5 does not constitute a transactive move and is not part of the cluster.

Our coding scheme was heavily informed by that originally developed by Goos et al.~\cite{Goos2002}, but is different in a few ways. Specifically, we made three minor changes to the original scheme when adapting it for use in our study. First, we  coded for different subtypes of new ideas (suggesting an approach, making an observation, etc.), whereas Goos et al. did not. Second, the original framework identified three types of transactive moves: self-disclosure, other-monitoring, and feedback requests. {Because students in our study occasionally asked each other for new explanations or suggestions, our scheme includes a fourth type of transactive move: idea requests}. Third and last, Goos et al. distinguished between transactive moves that are double-coded as metacognitive moves and those that are not (referred to as ``metacognitive transacts" and ``non-metacognitive transacts," respectively). While our scheme allows for a single utterance to be double-coded in this way, we do not distinguish between metacognitive and non-metacognitive transacts in our analysis. Our focus is on reciprocated verbal exchanges between two students, not individual conversational turns. To this end, our operational definitions of nodes and clusters are sufficient to capture metacognitive back-and-forth interactions.


\subsubsection{Cross-episode analysis: Emergent themes}

{Since there is limited research on SMM, we did not have any \emph{a priori} predictions for how such social interactions would regulate the troubleshooting activity.  Thus, after examining all 27 episodes and identifying a total of 23 clusters in student dialogue, we re-examined all of the clusters together to allow for the identification of common themes.  To accomplish this, we employed a grounded theory approach in order to characterize the broad nature of discussions that occurred in clusters. Grounded theory is a data-driven methodology in which data are categorized on the basis of emergent themes, and then refined into more inclusive groupings~\cite{Glaser1967,Bryant2007}.  Categorization was initially performed by the first author, verified by the second author, and was discussed by all project collaborators.}

{We identified two nonoverlapping categories of cluster:}
\begin{itemize}
\item[] {\textbf{Collective strategizing:} During each of these clusters, one or more approaches were critiqued, refined, and/or enacted.}
\item[] {\textbf{Shared understanding:} During each of these clusters, both students agreed to accept or reject a prediction, explanation, or interpretation of an observation in response to an other-monitoring move.}
\end{itemize}
{Most clusters fit into one of these two categories, but some did not. For example, in one cluster, students were working together to interpret a confusing figure in the datasheet; since the students were not reasoning about the circuit itself, this cluster did not fit into either category. In the other cases, students were discussing explanations or predictions, but did not reach consensus on any ideas; since there was no consensus, these clusters did not fit into the shared understanding category.}


\section{Results}\label{sec:results}

We describe data and findings from two different qualitative analyses of students' socially mediated metacognition. First, we provide an overview of students' metacognitive behaviors within each category of episode. {Then, we look for patterns among clusters across all types of episodes.} Throughout our discussion, we refer to groups of students using letters A to H. Within a group, individuals are labeled A1 and A2, B1 and B2, and so on. After presenting transcripts of dialogue between two students, we directly map the dialogue to the framework for SMM described in in Sec.~\ref{sec:within-episode}. In doing so, we aim to address whether and how students engage in SMM while troubleshooting an electric circuit, which is the major focus of our research.


\tabA

\subsection{Results from within-episode analyses}

The results of coding for metacognitive moves, transactive moves, and clusters are summarized in Tables~\ref{tab:metacog}, \ref{tab:transact}, and \ref{tab:clusters}, respectively. Based on these results, several patterns can be discerned. For example, {students suggested at least one approach in every episode, but assessed them in relatively few episodes (Table~\ref{tab:metacog}). In addition, clarification was a common type of transactive move: it was observed at least once in all-but-one episode.} Meanwhile, students requested feedback from one another more frequently than they asked one another for new ideas (Table~\ref{tab:transact}). Discrepant output episodes yielded not only the largest diversity of metacognitive and transactive moves, but also the highest frequency of clusters across groups (Table~\ref{tab:clusters}).

During IS, SH, and RD episodes, some groups did not engage in dialogue that contained a cluster. In all but one such cases, no clusters were observed because the dialogue contained only non-overlapping nodes. Lack of node overlap was due to one of two patterns: students changing the topic of conversation between nodes, or successive conversational turns that were either both metacogntive or both transactive. In one case where no clusters were observed---the IS episode for group B---one student dominated the conversation. The non-dominant speaker was  actively listening (e.g., by saying, ``Okay," ``Mhm," ``Yes," ``Right," and so on) rather than contributing to the conversation in metacognitive or transactive ways. Group B's IS episode was the only episode in our data set that contained no nodes; nodes were present in all other episodes.  {In all episodes in which no clusters were observed, students were not metacognitively engaging in each other's ideas.}

Next, we discuss and further characterize all four categories of episodes to better illuminate how students engaged in socially mediated metacognition. We limit our discussion to excerpts of dialogue that contain clusters, {as they best capture instances of social mediation of metacognition}. Information added to the transcripts for clarity is indicated by square brackets.

\tabB


\subsubsection{Initial strategizing episode}

Initial strategizing episodes were identified in all eight interviews. Each IS episode consisted of the first one or two minutes of the troubleshooting activity, starting just after students finished receiving instructions from the interviewer and ending when they began either making measurements or carrying out a detailed inspection of the circuit. During these episodes, students were transitioning from formulating the problem description to performing initial diagnostic tests. {The nature of this transition is consistent with the observed subtypes of metacognitive moves (Table~\ref{tab:metacog})}. Making observations, stating facts, and suggesting approaches were present in most or all IS episodes. Meanwhile, all subtypes of assessment were relatively infrequent. No students suggested explanations---potentially because students did not yet have information about the circuit's performance and therefore had little to explain.

As can be seen in Table~\ref{tab:clusters}, clusters were observed in three groups during IS episodes. The low frequency of IS clusters may be a reflection of the relative lack of information about the circuit's performance at the start of the activity compared to other episodes, which took place after the students had performed diagnostic measurements. IS clusters focused on formulating the problem description and prioritizing future measurements.

The following exchange between E1 and E2 is an example of an IS cluster that focused on forming an initial understanding the circuit's performance:\\
\vspace{6pt}\\
\noindent
\begin{tabularx}{\columnwidth}{*{4}{l}X}
\cline{2-2}
\multicolumn{1}{l|}{1.} & \multirow{2}{*}{A} & &
E1: & Alright. Cool. Well, how do you want to start this out? We could work out theoretically what it should do to start.\\ \cline{3-3} 
\multicolumn{1}{l|}{2.} & \multicolumn{1}{l|}{} & \multirow{2}{*}{B} &
E2: & They give us a pretty good transfer function right there [on the schematic].\\ \cline{2-2}
& \multicolumn{1}{l|}{} & & & \emph{E1 looks at the handout.} \\
3. & \multicolumn{1}{l|}{} & &
E1: & Okay. Cool. That makes sense, just like inverting and not inverting smashed together.\\ \cline{3-3}
\end{tabularx}
\vspace{6pt}\\
This exchange took place immediately after the interviewer finished introducing the task. Here, E1 initiated the conversation by asking a question about how to proceed {(1; idea request)} and suggesting an approach (1; new idea). E2 then remarked that the schematic included relevant information (2; new idea). Last, E1 stated that the schematic made sense to him (3; assessment) and elaborated his understanding of the circuit subsystems (3; self-disclosure). Through this discussion, the students supported each other in using the schematic to develop a model of the circuit as consisting of two distinct stages. Although this model did not inform further discussion during the initial strategizing episode, E1 and E2 later employed a split-half strategy, which relies on the identification of independently testable stages.

\tabC


\subsubsection{Discrepant output episode}

Discrepant output episodes were identified in all eight interviews. Each DO episode consisted of the discussions that followed immediately after students observed that the output of the circuit was a constant dc voltage, which did not match their expectations. {As can be seen in Tables~\ref{tab:metacog} and \ref{tab:transact}, every subtype of transactive and metacognitive moves was present in at least one DO episode}. All groups engaged in assessing results, suggesting strategies, and monitoring ideas; approach assessments were observed in half of DO episodes, compared to one or none in other episodes. Such conversational moves are consistent with the cognitive task transition that defines DO episodes: students transitioned from generating causes to performing additional diagnostic tests.

Most groups carried out actions that would further their understanding of the malfunctioning circuit. Some, however, did not appear to use the information gained from their observations to inform and constrain the investigations immediately following the episode. Specifically, {two groups} tested the signal with an ac input, but subsequently decided to measure resistor values. These groups did not consider that a problem with resistor values could not fully account for the faulty dc output signal they had observed. Similarly, {one group} made a decision to re-investigate the circuit, but this decision was not attached to a specific hypothesis as to how their course of action would help advance their understanding.

Clusters were observed in each DO episode (Table~\ref{tab:clusters}), and we highlight two examples here. In each of these examples, students' metacognitive discussions directly informed the their subsequent investigations of the malfunctioning circuit.

Students C1 and C2 used an ac input voltage to test the circuit. To monitor the output signal, they used two separate cables that connected the output of stage 2 to two different channels of the oscilloscope. This excerpt begins just as the students observed the output signal for the first time:\\
\vspace{6pt}\\
\noindent
\begin{tabularx}{\columnwidth}{*{5}{l}X}
\cline{2-2}
\multicolumn{1}{l|}{1.} & \multirow{2}{*}{A} & & &
C1: & That's getting us a dc voltage. Or is that oscillating? That's bizarre. Why is it---? \\ \cline{3-3}
\multicolumn{1}{l|}{2.} & \multicolumn{1}{l|}{} & \multirow{2}{*}{B} & &
C2: & Yep, these guys [both channels] are measuring the same dc. \\ \cline{2-2} \cline{4-4}
3. & \multicolumn{1}{l|}{} & \multicolumn{1}{l|}{} &\multirow{2}{*}{C} &
C1: & Is something just being a voltage divider or something? What's that value? \\ \cline{3-3}
& & \multicolumn{1}{l|}{} & &
& \emph{C1 and C2 adjust oscilloscope settings.}\\ 
4. &  & \multicolumn{1}{l|}{} & &
C2: & Fourteen volts. \\  \cline{2-2} \cline{4-4}
\multicolumn{1}{l|}{5.} & \multirow{2}{*}{D}  & & &
C1: & It's probably saturated. \\ 
\multicolumn{1}{l|}{6.} &  & & &
C2: & No, if it was saturated it would still oscillate, right? It would just clip at the sides? So, I mean, more likely that 14 is pretty close to this guy [the power supply]. Maybe one of the [breadboard] rails is bad underneath. That's certainly possible. \\ \cline{2-2}
\end{tabularx}
\vspace{6pt}\\
In this exchange, C1 described the output of the circuit (1; new idea), called it ``bizarre" (1; assessment), and questioned whether it was oscillating {(1; feedback request)}. C2 then confirmed C1's initial description of the dc output (2; other-monitoring). Next, C1 questioned if this could have been the result of voltage division {(3; feedback request)} and suggested monitoring the value of the output (3; new idea). After engaging in this cluster, the students adjusted the oscilloscope settings to better read the  signal, and C2 verbalized the value of the output voltage (4; new idea). C1 suggested saturation as a potential explanation for this dc value (5; new idea). In response, C2 critiqued C1's explanation (6; other-monitoring) and clarified his criticism by describing the typical waveform of a saturated signal (6; self-disclosure). C2 went on to note the similar magnitude of the output signal and the power supply (6; new idea), ultimately suggesting an alternative explanation for the output (6; new idea). Nodes A to C form a cluster focused on the characteristics of the output signal. {Because turns 4 and 5 are both metacognitive moves, node D is separate from the cluster.} In this example, C2 monitored the explanatory power of his partner's explanation and rejected an erroneous hypothesis about the circuit's performance.

Students F1 and F2 also used an ac input voltage to test the circuit. This excerpt begins just as the the students were discussing what to do next, after having observed faulty behavior of the circuit:\\
\vspace{6pt}\\
\noindent
\begin{tabularx}{\columnwidth}{*{7}{l}X}
\cline{2-2}
\multicolumn{1}{l|}{1.} & \multirow{2}{*}{A} & &  & & &
F1: & Should we---? We should make sure that this [the inverting input of op-amp 1] is zero volts. \\ \cline{3-3}
\multicolumn{1}{l|}{2.} & \multicolumn{1}{l|}{} & \multirow{2}{*}{B} & & & &
F2: & Um, this should not be zero volts. It should be the same as $\vin$, I think. Right? It should be zero down here [at ground]. \\ \cline{2-2} \cline{4-4}
3. & \multicolumn{1}{l|}{} & \multicolumn{1}{l|}{} & \multirow{2}{*}{C} & & &
F1: & Okay. But. Where is that coming from? The feedback or something? \\ \cline{3-3} \cline{5-5}
4. & & \multicolumn{1}{l|}{} & \multicolumn{1}{l|}{} & \multirow{2}{*}{D} & &
F2: & It's just the golden rule of the op-amp that the inputs wanna be the same. \\  \cline{4-4} \cline{6-6}
5. &   & &  \multicolumn{1}{l|}{} & \multicolumn{1}{l|}{} &  \multirow{2}{*}{E}  &
F1: & Yeah, but how could the negative terminal be the same as the positive terminal at all times? \\ \cline{5-5}
6. &  & & & \multicolumn{1}{l|}{} & &
F2: & I don't know how it works. \\ \cline{6-6}
\end{tabularx}
\vspace{6pt}\\
Here, F1 suggested that he and his partner perform a particular diagnostic measurement (1; new idea). F2 disagreed with an implicit assumption in F1's suggestion (2; other-monitoring), and articulated alternative predictions about the circuit's expected performance (2; new idea). In response, F1 asked F2 for more information about his ideas (3; other-monitoring), and asked if feedback might be a relevant mechanism {(3; feedback request)}. To justify his reasoning (4; self-disclosure), F2 recited one of the op-amp golden rules (4; new idea). Next F1 asked F2 for further explanation (5; other-monitoring), which F2 said he could not provide (6; assessment). Nodes A to E are all part of the same cluster. This excerpt highlights how students F1 and F2 explored the limitations of their own knowledge while they were simultaneously drawing upon that same knowledge to form predictions. Despite not having a complete explanation for the ideal behavior of an op-amp, they were able to use the golden rules to make concrete predictions later in the troubleshooting task.


\subsubsection{Split-half episode}

Split-half episodes were identified in five interviews. Each SH episode consisted of the discussions that followed immediately after students successfully employed a split-half strategy, and ended when they began a new set of measurements. During SH episodes, students were transitioning from generating hypotheses (namely, the hypothesis that stage 1 was functional and hence any faults resided in stage 2) to performing additional tests. As can be seen in Tables~\ref{tab:metacog} and \ref{tab:transact}, several subtypes of metacognitive and transactive moves were present in all five SH episodes: suggesting approaches, articulating facts, clarifying one's own ideas, and monitoring another's ideas. Clusters were present in four of the five episodes (Table~\ref{tab:clusters}). In this section, we discuss a single SH episode in its entirety, noting that this episode was representative of most episodes within this category.

The episode we discuss begins immediately after students G1 and G2 agreed that stage 1 was functioning as expected (asterisks indicate simultaneous speech):\\
\vspace{6pt}\\
\noindent
\begin{tabularx}{\columnwidth}{*{5}{l}X}
1. & & &  &
G1: & So, we can isolate this part. \\ \cline{2-2}
\multicolumn{1}{l|}{2.} & \multirow{2}{*}{A} & & &
G2: & So then this op-amp, so then, ahh let's see. This right here [the inverting input] should be ground. \\ \cline{3-3}
\multicolumn{1}{l|}{3a.} & \multicolumn{1}{l|}{} & \multirow{2}{*}{B} & &
G1: & Yeah, yeah, this is virtual ground--- \\ 
\multicolumn{1}{l|}{*}  & \multicolumn{1}{l|}{} & & &
G2: & Virtual ground. \\  
\multicolumn{1}{l|}{3b.} & \multicolumn{1}{l|}{} & &  & 
G1: & ---right here. No current's going through here [into the inverting input]. So, from there, we can say current through here [$R_3$] is equal to current through there [$R_4$]. \\ \cline{2-2}
%
\cline{4-4}
%
%
%
4. & \multicolumn{1}{l|}{\textcolor{white}{A}}  & \multicolumn{1}{l|}{\textcolor{white}{B}}   &  \multirow{2}{*}{C} &  
G2: & So this resistor right here, the $R_3$, that should have a drop of 10 volts then. Because you have ground right here [at the inverting input]. \\ \cline{3-3}
5a. &  & \multicolumn{1}{l|}{}   & & 
G1: & Yeah, yeah, you're right, because this [the inverting input] is zero volts, this [stage 1's output] is 10 volts, so we should be losing--- \\
* &  & \multicolumn{1}{l|}{} & & 
G2: & Ten volts across there. \\ 
5b. &  & \multicolumn{1}{l|}{}  & & 
G1: & ---10 volts across that resistor. Okay so, I'll look at, we should be losing 10 volts across here. Alright so let's, let's check it out.\\ \cline{4-4}
\end{tabularx}
\vspace{6pt}\\
In the above exchange, we treat G1's speech in turns 3a to 3b and 5a to 5b as a continuous conversational turn despite G2's simultaneous speech. This SH episode started when G1 suggested that stage 2 could be isolated (1; new idea). G2 then examined the circuit and articulated a prediction about the voltage of the inverting input of stage 1 (2; new idea); this prediction is consistent with the first golden rule for op-amps. G1 endorsed G2's idea (3a; other-monitoring) and further clarified that the input would be a ``virtual" ground, which in this context indicates that it is not directly connected to ground (3b; self-disclosure). Next, G1 articulated a prediction that the currents through resistors $R_3$ and $R_4$ would be equal (3b; new idea); this prediction is consistent with the second golden rule for op-amps. G2 subsequently predicted the voltage drop across $R_3$ (4; new idea), and justified his prediction (4; self-disclosure). Last, G1 endorsed and built upon G2's prediction (5a; other-monitoring), ultimately suggesting that they perform a particular measurement to test the prediction (5b; new idea).

Nodes A to C form a cluster. In this cluster, students combined their knowledge of the golden rules for op-amps and the output of stage 1 to make a testable prediction for the voltage across resistor $R_3$.  Similar exchanges were documented and analyzed in three of the other four groups that employed a split-half strategy. The only outlier was group D. Students D1 and D2 began retesting the voltages in stage 2 without making new predictions about the circuit's expected performance. We note that group D was ultimately unsuccessful in repairing the circuit within the time constraints of the interview.


\subsubsection{Replacement decision episode}

Replacement decision episodes were identified in seven interviews. RD episodes focused on the decision to replace the op-amp in stage 2. The episodes began when students started discussing the last measurement made prior to the replacement, and ended when students began to replace the op-amp. Every group who replaced the op-amp had previously replaced resistor $R_3$. RD episodes constituted a transition from performing tests to repairing and evaluating the circuit. Each group that successfully replaced the op-amp considered, yet subsequently rejected, problems occurring elsewhere in the circuit. We discuss a single excerpt from group C that highlights the collaborative establishment and justification of the group's decision to replace the op-amp.

Earlier in the interview, students C1 and C2 erroneously replaced the first op-amp. Just prior to the RD episode, they re-measured the input signal and the outputs of both stages. They noted that the first stage functioned as expected, but the output of stage 2 was still a large dc value. The excerpt below begins immediately after the students measured the inputs to second op-amp:\\
\vspace{6pt}\\
\noindent
\begin{tabularx}{\columnwidth}{*{5}{l}X}
\cline{2-2}
\multicolumn{1}{l|}{1a.} & \multirow{2}{*}{A} & &  &
C2: & Pin three [of the second op-amp] is--- \\
\multicolumn{1}{l|}{*} & & & &
C1: & Zero. \\ 
\multicolumn{1}{l|}{1b.} & & & &
C2: &---in fact zero. However pin two [of the second op-amp] is not zero, right? And that's the problem. That's the op-amp.\\ \cline{3-3} 
\multicolumn{1}{l|}{2.} & \multicolumn{1}{l|}{}  & \multirow{2}{*}{B}  &&
C1: & So that's saying that we're losing our--- The op-amp is wrong, too? \\  \cline{2-2} \cline{4-4}
3. &   \multicolumn{1}{l|}{}  &  \multicolumn{1}{l|}{} & \multirow{2}{*}{C} & 
C2: & Yeah, it must be. That means the golden---I mean, the first one could've been fine, in retrospect---but certainly the second one is not working, because the golden rules are not being followed here. \\ \cline{3-3}
4. &  & \multicolumn{1}{l|}{}  & &
C1: & Okay, that's not it. Want to switch that guy out?\\ \cline{4-4}
5. &  & & &
C2: & Yeah. \\ 
\end{tabularx}
\vspace{6pt}\\
Here, we treat C2's speech in turns 1a and 1b as a continuous conversational turn. This exchange began when C2 observed the input voltages of the op-amp in stage 2 (1b; new idea), noting that there was a problem since the inputs had different values (1b; assessment). This assessment is consistent with the first golden rule for op-amps. C1 then asked whether this meant there was a problem with the op-amp in stage 2 {(2; feedback request)}. In response, C2 endorsed C1's idea (3; other-monitoring), suggested they may have misdiagnosed the op-amp in stage 1 (3; assessment). C2 also justified C1's tentative hypothesis about the op-amp in stage 2 by referencing the golden rules for op-amps (3; self-disclosure). In turn 4, it is unclear to the authors what C1 was referencing when he said, ``Okay, that's not it." However, he went on to suggest a repair, namely, the replacement of the op-amp in stage 2 (4; new idea). This suggested approach was taken up by C2.

In this excerpt, students C1 and C2 made sense of a new set of voltage measurements, with some confirming, but others superseding their earlier work. They used their results to justify replacing the second op-amp, and to reflect upon their earlier misdiagnosis of the first op-amp. Including group C, six of the seven groups who successfully repaired the circuit justified their decision to replace the op-amp in stage 2 by synthesizing information from their most recent measurements and those performed throughout the interview.


\subsection{Results from cross-episode analysis}

{In addition to identifying and describing examples of SMM in each episode category, we also looked for conversational themes among clusters from all groups and all episodes. We organized clusters into two separate categories: (i) collective strategizing about the troubleshooting process, and (ii) shared understanding of the circuit's behavior. These emergent categories are consistent with the high rates of metacognitive moves in which new approaches were suggested (Table~\ref{tab:metacog}) and transactive moves focused on monitoring or clarifying ideas (Table~\ref{tab:transact}). Out of 23 total clusters in our dataset, 8 focused on collective strategizing, 11 on shared understanding, and 4 fit into neither category. A breakdown of the number of groups in which clusters of either kind were observed is provided in Table~\ref{tab:clusters}. In this section, we present one example of a cluster from each category.}


\subsubsection{Collective strategizing}

Clusters about {collective strategizing} mostly occurred during the first half of the troubleshooting process, within initial strategizing and discrepant output episodes (Table~\ref{tab:clusters}). In these episodes, students were formulating the problem description via visual inspection of the circuit (IS episodes) and reacting to the first measurement of the malfunctioning circuit output (DO episodes). Both episodes involved transitions to the troubleshooting subtask of performing tests. Thus, rich metacognitive dialogue about approaches for repairing the circuit coincided with students' early formative and diagnostic observations.

As an example, we present a cluster from a DO episode. In this excerpt, the students in group E had just observed that the output of the circuit was a constant dc voltage, and they began the process of deciding how to proceed in repairing the circuit:\\
\vspace{6pt}\\
\noindent
\begin{tabularx}{\columnwidth}{*{7}{l}X}
\cline{2-2}
\multicolumn{1}{l|}{1.} & \multirow{2}{*}{A} & &  & & &
E2: & Do we even check if these are the right chips? That would be kind of stupid. \\ \cline{3-3}
\multicolumn{1}{l|}{2.} & \multicolumn{1}{l|}{} & \multirow{2}{*}{B} & & & &
E1: & It would probably be a good call. \\ \cline{2-2} \cline{4-4}
3. & \multicolumn{1}{l|}{} & \multicolumn{1}{l|}{} & \multirow{2}{*}{C} & & &
E2: & Okay, I guess we do have--- Can we just, like, pull that chip out and replace it?\\ \cline{3-3} \cline{5-5}
4. & & \multicolumn{1}{l|}{} & \multicolumn{1}{l|}{} & \multirow{2}{*}{D} & &
E1: & Yeah, I mean, it will be like the brute force method of making sure it's the right chip. Pull it out and put the right one in. \\  \cline{4-4} \cline{6-6}
5. &   & &  \multicolumn{1}{l|}{} & \multicolumn{1}{l|}{} &  \multirow{2}{*}{E}  &
E2: & What we could do is get out a probe and we can just go through the first one and measure $\vout$, and we could see if that's what we expect it to be. \\ \cline{5-5}
6. &  & & & \multicolumn{1}{l|}{} & &
E1: & Yeah, for sure. And then we'll measure all the power to make sure it's doing what it should be doing. \\ \cline{6-6}
\end{tabularx}
\vspace{6pt}\\
Here, E2's suggestions in turns 1 and 3 were phrased as questions. The exchange began when E2 suggested a potential strategy for troubleshooting the circuit (1; new idea). E1 affirmed that the strategy could be productive (2; assessment, other-monitoring). E2 then suggested a new, related strategy (3; new idea), which E1 called a ``brute force method" (4; assessment, other-monitoring). We note that ``brute force method" has a negative connotation in physics problem solving; it is often used to refer to an inelegant approach. In response, E2 suggested yet another approach for testing the circuit's performance (5; new idea). Based on the context, $\vout$ refers to the output of stage~1 in turn 5. E1 endorsed and built upon this idea (6; other-monitoring) by suggesting different, additional tests (6; new idea).

Together, nodes A to E form a single cluster in which the students proposed and evaluated four different approaches: checking if the chips were correct (turn 1), replacing a chip (turn 3), measuring the output of stage~1 (turn 5), and measuring the voltage of the power rails (turn 6). The suggestions related to checking or replacing the op-amp chips were discarded, and the students began measuring voltages after this exchange.


\subsubsection{Shared understanding}

Clusters about shared understanding mostly occurred after the initial strategizing episode, i.e., during the discrepant output, split-half, and replacement decision episodes (Table~\ref{tab:clusters}). In these episodes, students were generating causal hypotheses about the source of malfunction in the circuit (DO and SH episodes) and proposing a potential repair (RD episodes). Thus, {reaching consensus on predictions, explanations, and interpretations of observations} through back-and-forth metacognitive exchanges occurred when one student was unsure of what claims were being made by a partner, or when both students were working together to understand the actual performance of the circuit.

As an example, we present a cluster from an RD episode. In this excerpt, the students in group A were interpreting a measurement of the negative power rail:\\
\vspace{6pt}\\
\noindent
\begin{tabularx}{\columnwidth}{*{4}{l}X}
\cline{2-2}
\multicolumn{1}{l|}{1.} & \multirow{2}{*}{A} & &
A1: & And that's at 15 and a half.\\ \cline{3-3} 
\multicolumn{1}{l|}{2a.} & \multicolumn{1}{l|}{} & \multirow{2}{*}{B} &
A2: & That's at plus 15 and a half? Oh, did you measure it backwards?\\ 
\multicolumn{1}{l|}{*} & \multicolumn{1}{l|}{} & &
A1: & Yeah.\\
\multicolumn{1}{l|}{2b.} & \multicolumn{1}{l|}{} & &
A2: & Did you have the leads flipped?\\ \cline{2-2}
3. & \multicolumn{1}{l|}{} & &
A1: & Yeah, yeah, yeah. That's fine.\\ \cline{3-3}
\end{tabularx}
\vspace{6pt}\\
At the beginning of this excerpt, A1 verbalized the measured value of the negative rail voltage (1; new idea). A2 questioned the reported value (2a; other-monitoring), asked if his partner had measured the voltage backwards (2a; other-monitoring), and clarified what he meant by ``backwards" (2b; self-disclosure). A1 affirmed that he did attach the leads of the multimeter backwards and that the measurement was in fact consistent with expectations (3; assessment).

In this interaction, nodes A and B form a cluster during which the students collaboratively clarified that A1's measurement was not the result of an actual flaw in the circuit, but rather stemmed from an incorrect measurement procedure. After the exchange, A2 began inspecting the connections of the circuit to ensure that it was constructed properly, indicating that he no longer questioned the measurement{; hence, A1 and A2 were in agreement about the interpretation of the original measurement}.


\section{Discussion and Limitations}\label{sec:discussion}

Our results provide insight into whether and how students engage in socially mediated metacognition while troubleshooting a malfunctioning op-amp circuit. Here, we focus on two major findings of this work, each corresponding to one of our research questions. First, in our study, students did indeed engage in SMM when troubleshooting a circuit (RQ1). {Second, reciprocated metacognitive dialogue (i.e., clusters) arose when students were {collectively strategizing about which measurements to perform, or reaching a shared understanding of the circuit's behavior}  (RQ2).} In addition to elaborating upon these findings, we draw on relevant studies to help contextualize our work and identify areas for potential future investigation.

{We observed multiple groups engaging in SMM in each of four strategic and evaluative episodes during the troubleshooting process:} (i) developing initial troubleshooting strategies, (ii) observing the discrepant output of the circuit for the first time, (iii) employing the split-half strategy to isolate the source of malfunction to one part of the circuit, and (iv) deciding to replace a faulty component. Multiple examples of metacognitive moves, transactive moves, nodes, and clusters were observed during peer interactions among all eight pairs of students in our study. Clusters occurred most frequently after students made measurements of the malfunctioning behavior of the whole circuit (discrepant output episodes) or the functional behavior of the first subsystem (split-half episodes). In these episodes, students were drawing on the new information provided by their measurements to generate causal hypotheses about the circuit's performance; then, based on these new ideas about the circuit, they were deciding which tests to perform.

By focusing on students' metacognitive discussions during transitions from one cognitive troubleshooting subtask to another, we were further able to gain insight into the role of SMM in repairing the circuit. Across all four categories of episodes, we observed that back-and-forth metacognitive exchanges facilitated troubleshooting in two major ways. First, students engaged in SMM when jointly deciding upon which troubleshooting approaches to employ. These decisions involved collaborative formation of hypotheses, predictions, and strategies for testing the circuit. Second, students engaged in SMM when trying to understand or refute each other's insufficiently substantiated ideas or incomplete analyses. In both cases, SMM was coupled to students' recognition that greater clarity was needed in order to know how to proceed with investigating the circuit. Such realizations prompted students to revisit each other's reasoning, refute erroneous ideas, and endorse productive suggestions---an inherently social metacognitive process.

When interpreting these findings, it is important to keep in mind two major limitations of our study. First, our participant pool was small and homogenous: of 16 students, most were white men, all had completed similar electronics lab courses, and all were enrolled in selective, predominantly white, research-intensive universities. {Therefore, additional studies with more diverse populations may identify different metacognitive social dynamics that arise during students' collaborative troubleshooting of malfunctioning apparatuses.} Second, the theoretical foundations of our study focus primarily on cognitive and metacognitive social dynamics. However, other work has emphasized that troubleshooting is a frustrating task that requires perseverance, creativity, confidence, patience, and a belief that troubleshooting is a normal part of physics experiments~\cite{Dounas-Frazer2017,Dounas-Frazer2016b,Estrada2012,MacPherson1998}. In this sense, our study does not fully capture the troubleshooting experience. With these limitations in mind, we identify implications for research and instruction.

In a previous study~\cite{Dounas-Frazer2016a}, we used data from the participants in the present study to investigate whether and how they used model-based reasoning when troubleshooting an electric circuit. We found that students ``engaged in multiple, distinct iterations of model-based reasoning while navigating the cognitive" subtasks of the troubleshooting process~(p.~18), and we argued that students' ability to troubleshoot and their ability to model physical systems are complementary experimental physics skills. {Given that socially mediated metacognition arises when students troubleshoot}, it is likely that metacognitive dialogue arises when  work together to construct and refine models and apparatuses in contexts other than repairing a malfunctioning system. Future work could explore the theoretical and empirical connections between the SMM framework and frameworks for model-based reasoning.

The social environment of many instructional settings is not only due to interactions among students, but also to those between students and instructors. Indeed, Goos et al.~\cite{Goos2002} argued that ``the teacher has a crucial role to play in orchestrating fruitful collaboration," by, for example, scaffolding ``students' selection of strategies, identification of errors, and evaluation of answers" (p.~220). Along these lines, Dounas-Frazer and Lewandowski~\cite{Dounas-Frazer2017} found that electronics lab instructors' self-reported practices for teaching students how to troubleshoot align well with the cognitive apprenticeship paradigm of instruction: asking students to articulate their own understanding, coaching students about different troubleshooting strategies, and/or modeling their (instructors') own approaches to troubleshooting by verbalizing their thought processes while repairing a circuit in front of student observers who watch and listen. Although these interactions occur between students and instructors rather than among student groups, they are nevertheless examples of social metacognitive dynamics. Hence, the SMM framework could be a useful tool for characterizing and evaluating instructors' teaching practices in electronics and other lab courses.

Finally, we note that teaching students how to troubleshoot circuits may benefit from explicit classroom norms about collaboration---especially in lab courses that require students to work in groups. Cognitive apprenticeship teaching practices, which are well suited to developing students' competence with cognitive aspects of troubleshooting, could be supplemented by deliberate efforts to support students' metacognitive regulation of their lab partners' thinking. For example, lab instructors could encourage students to ask themselves \emph{and each other}, ``What are you doing, why are you doing it, and how does it help?"~(cf. Schoenfeld~\cite{Schoenfeld1992}).



\section{Summary}\label{sec:conclusion}

We developed a troubleshooting activity in which students attempted to diagnose and repair a malfunctioning op-amp circuit. Audiovisual data were collected for eight pairs of students from two separate institutions. We analyzed transcripts of student dialogue using an \emph{a priori} framework for socially mediated metacognition and an emergent thematic analysis of clusters, {a form of reciprocated peer-to-peer metacognitive regulation}.

Our findings demonstrate a good mapping between students' performance of an experimental physics task and the SMM framework, which was originally developed by Goos et al.~\cite{Goos2002} in the context of high school students solving physics-based math problems. In addition, our findings align well with the recommendations of Lippmann Kung and Linder~\cite{LippmannKung2007}, who stressed the importance of documenting not just whether students engage in metacognition, but how their metacognition informs their subsequent actions when working on physics lab activities. We have shown how the SMM framework can be coupled with other frameworks (in this case, a cognitive task analysis of troubleshooting) to provide a rich picture of students' reactions to metacognitive dialogue: which claims are accepted, which strategies are adopted, which measurements are performed, and how those claims, strategies, and measurements facilitate transitions between different phases of problem solving. This suggests that the SMM framework can be a productive tool for analyzing other types of collaborative experimental physics problem solving.


\acknowledgments 

The authors would like to acknowledge the contributions made by members of the UMaine Physics Education Research Laboratory for their input on framing and interpreting some of these results. {Benjamin Pollard and Jessica Hoehn provided useful feedback on the manuscript.} This material is based upon work supported by the National Science Foundation under Grant Nos. DUE-1245313, DUE-1323101, DUE-1323426, PHY-1125844, and DUE-0962805.

\end{document}